# Crystallization Behavior of ZBLAN Glass Under Combined Thermal and Vibrational Effects: Part II - COMSOL Simulation and Apparatus Redesign

Ayush Subedi[1*], Anthony Torres[1], Jeff Ganley[2]

[1]Materials, Science, Engineering, and Commercialization (MSEC), Texas State University, San Marcos, Tx 78666, United States

[2]Air Force Research Labs, Space Vehicles Directorate

*Corresponding author ayush.rajsubedi@gmail.com

**Abstract**

In Part I of this study, vibration-assisted heat treatments of ZBLAN ($ZrF_4$-$BaF_2$-$LaF_3$-$AlF_3$-$NaF$) glass revealed irregular crystallization at higher vibration levels, attributed to intermittent loss of thermal contact between the sample and the inner silica ampoule wall. The present work (Part II) investigates this mechanism through finite element modeling (FEM) and experimental validation. COMSOL Multiphysics simulations incorporating conduction, radiation, and contact resistance confirm that intermittent contact markedly reduces heat transfer efficiency, lowering the sample temperature. To mitigate this effect, the experimental setup was redesigned with a four-degree inclination to maintain stable contact during vibration. Subsequent experiments at vibration levels H3-H5 demonstrated uniform heating and consistent crystallization behavior.

Comprehensive microscopic, Scanning Electron Microscopy (SEM), Energy-Dispersive X-ray Spectroscopy (EDS), and Atomic Force Microscopy (AFM) analyses revealed that even at subtle vibration levels (~50 Hz), partially crystallized ZBLAN transformed into well-developed crystalline structures near 360 °C. With increasing vibration amplitude, amorphous ZBLAN began forming incipient crystalline phases around 330 °C, and at higher frequencies (~100 Hz), partial crystallization initiated at approximately 350 °C. These results indicate that higher vibration frequencies accelerate nucleation, enhance heat transfer, and reduce the effective fiber-drawing temperature window by about 30 °C. Prolonged exposure above 330 °C under vibration promotes unwanted phase transitions, emphasizing the need for precise thermal and vibrational control. This study establishes a predictive framework for vibration-resistant ZBLAN processing applicable to both terrestrial and microgravity environments.

## 1. Introduction

ZBLAN ($ZrF_4$-$BaF_2$-$LaF_3$-$AlF_3$-$NaF$) glass is widely regarded as one of the most promising materials for infrared optical fibers owing to its exceptionally broad transmission window, low intrinsic attenuation, and chemical stability [1]. These attributes make it an attractive candidate for high-performance communication and sensing applications across the near- to mid-infrared range. However, large-scale terrestrial fabrication of high-quality ZBLAN fibers remains challenging because of the glass's strong tendency to crystallize during processing, which introduces scattering centers and significantly increases optical losses [2]. While microgravity environments have demonstrated the ability to suppress such devitrification by minimizing buoyancy-driven convection [3], [4] and sedimentation, reproducing similar stability under normal gravity (1g) conditions demands a deeper understanding of how temperature gradients, vibration, and local heat-transfer dynamics collectively influence nucleation and crystallization.

In Part I of this study [5], vibration assisted heat treatment experiments demonstrated that mechanical vibration could accelerate nucleation and lower the crystallization onset temperature, indicating enhanced atomic mobility. However, several samples treated at higher vibration levels (H3, H4, and H5) exhibited irregular crystallization behavior, showing minimal morphological changes despite being exposed to critical crystallization temperatures [5]. These inconsistencies were attributed to a jostling effect, where excessive vibration caused the ZBLAN samples to intermittently lose contact with the inner silica ampoule wall. This mechanical decoupling reduced conductive heat transfer efficiency and prevented the samples from achieving the target temperature necessary for crystallization within the short processing time.

The present Part II investigates this hypothesis using finite element modeling (FEM) in COMSOL Multiphysics to simulate transient heat transfer within the ZBLAN-silica heating system under vibration-induced displacement. The model incorporates conduction, radiation, and thermal contact resistance to replicate realistic experimental conditions and quantify temperature gradients arising from intermittent contact loss. The simulation results confirm that intermittent contact significantly limits heat transfer efficiency at higher vibration amplitudes. Guided by these findings, a redesigned apparatus was developed to ensure consistent thermal coupling and minimize sample movement during vibration.

Subsequent experiments were conducted using this improved setup at vibration levels H3, H4, and H5. The treated samples were then systematically characterized using optical microscopy, scanning electron microscopy (SEM), and energy-dispersive X-ray spectroscopy (EDS) to evaluate crystal morphology, surface features, and compositional variations. A detailed comparative analysis of multiple samples, particularly those exposed to higher vibration frequencies, is presented in this paper. These samples exhibited distinct crystallization patterns with densely populated crystalline regions, providing deeper insight into the correlation between vibration amplitude, heat transfer uniformity, and crystallization behavior. Collectively, this work integrates simulation, apparatus optimization, and advanced characterization to strengthen the understanding and control of vibration-assisted crystallization in ZBLAN, paving the way for reliable terrestrial and space-based fiber drawing.

## 2. Modeling and Simulation Methodology

To validate the hypothesis proposed in Part I [5] that intermittent jostling of the ZBLAN sample during vibration reduced effective thermal contact with the silica ampoule and consequently prevented sufficient heating for crystallization, finite element simulations were conducted using COMSOL Multiphysics (v6.2). The model employed the Heat Transfer in Solids physics interface, which simultaneously accounts for thermal conduction and surface to surface radiation. Convective heat transfer was excluded because the silica ampoule was maintained under high vacuum during the experiment, leaving conduction and radiation as the only active heat transfer mechanisms. The results of this simulation were used to evaluate how intermittent loss of thermal contact could cause parts of the sample to remain in the amorphous state rather than undergoing the expected transition to the crystalline phase.

## 2.1 Model Geometry

To replicate the experimental setup and validate the hypothesis proposed in Part I [5], two distinct models were developed in COMSOL Multiphysics. Both designs were constructed using the exact dimensions of the silica ampoule and the ZBLAN glass sample used in the actual experiment to ensure geometric and thermal consistency between simulation and experiment. This approach enabled direct evaluation of the heat transfer behavior under identical spatial configurations.

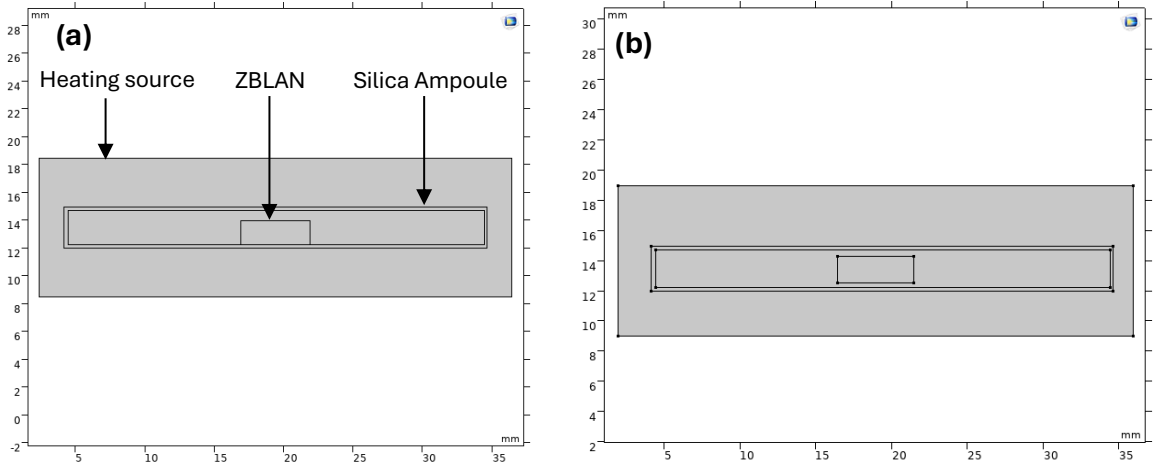

*Figure 1: COMSOL model geometries of (a) Design I and (b) Design II, illustrating the ZBLAN positioned inside the silica ampoule under contact and floating configuration respectively.*

As illustrated in Figure 1(a) and Figure 1(b), the simulation setup consists of multiple concentric regions. The outermost layer represents the heating source, which supplies heat uniformly through all four external walls. Inside the heater, a thin air layer separates the heating source from the silica ampoule, replicating the real experimental spacing. The ampoule itself encloses the ZBLAN glass piece, and the interior of the ampoule is maintained under vacuum conditions, eliminating any convective heat transfer within the cavity.

In Design I as represented by Figure 1(a), the ZBLAN sample is placed in direct contact with the inner wall of the silica ampoule, simulating stable contact conditions during vibration. Under this configuration, the ZBLAN receives heat through both thermal conduction at the contact interface and surface to surface radiation between the surrounding walls. This case represents ideal heating conditions where the sample remains thermally coupled to the ampoule wall throughout the process.

In Design II as represented by Figure 1(b), the ZBLAN sample is slightly displaced from the ampoule wall, leaving a narrow vacuum gap between the two surfaces. This design replicates the jostling behavior observed at higher vibration levels (H3-H5) in Part I [5]. Here, the conductive pathway is interrupted, and the sample receives heat solely through surface-to-surface radiation across the vacuum gap.

Both configurations were modeled as axisymmetric two-dimensional geometries to accurately represent the radial heat transfer behavior while maintaining computational efficiency. The outer surface of the heating layer was maintained at a constant temperature of 400 °C (673.15 K), replicating the experimental thermal conditions applied during testing. These models enable a direct quantitative comparison of temperature distribution and heat uptake between the full-contact and contact-loss scenarios, thereby elucidating the influence of intermittent jostling on thermal uniformity and providing numerical validation for the hypothesis proposed in Part I [5].

## 2.2 Governing Equations and Boundary Conditions

The heat transfer within the ZBLAN- silica ampoule system was modeled using the Heat Transfer in Solids interface of COMSOL Multiphysics (v6.2). This interface simultaneously accounts for thermal conduction and surface-to-surface radiation between participating surfaces. Since the silica ampoule was maintained under vacuum during all experimental trials, convective heat transfer was neglected. Thus, conduction and radiation were the only active heat transfer mechanisms within the modeled domain.

Following this setup, the transient temperature field within each solid domain was governed by the energy conservation equation for heat transfer in solids [6], [7]. This formulation captures the temporal evolution of temperature due to conduction within the materials and radiative exchange across interfaces. The governing equation can be expressed as,

$$\rho c_p \frac{\partial T}{\partial t} = \nabla \cdot (k \nabla T) + Q$$

where $\rho$ is the density, $c_p$ is the specific heat capacity, k is the thermal conductivity, and Q is the volumetric heat generation rate.

Under the vacuum conditions, the advection term is omitted because no fluid motion occurs, leaving conduction and radiation as the dominant heat transfer mechanisms.

At solid interfaces in direct contact (Design I), heat transfer follows Fourier's law [6], [7] of conduction:

$$q = -k \nabla T$$

While across the vacuum gap (Design II), radiative heat exchange between diffuse-gray surfaces is modeled by the viscosity-irradiation relation[6], [7]:

$$J = \epsilon \sigma T^4 + (1-\epsilon)G$$

$$q_{rad} = \epsilon \, (\sigma T^4 - G)$$

where J is the radiosity, G is the irradiation, ϵ is the emissivity and σ is the Stefan-Boltzmann constant.

The total irradiation G includes view-factor contributions from all participating surfaces and ambient surroundings, expressed as [6], [7]:

$$G_{amb} = F_{amb} \, \epsilon_{amb} \, \sigma T_{amb}^4$$

These coupled conductive and radiative formulations enable accurate simulation of transient heat transfer within the ZBLAN-silica-tungsten system.

The outer boundary of the heater was maintained at a fixed temperature of 400 °C, representing the experimentally applied heating condition. All other exterior boundaries were assumed adiabatic, preventing heat loss from the model's periphery.

At the contact interface between ZBLAN and the inner ampoule wall (Design I), perfect thermal contact was assumed, allowing continuous conduction across the interface. In contrast, for the floating configuration (Design II), the gap between the ZBLAN and ampoule wall was modeled as a vacuum region where heat transfer occurs solely via surface-to-surface radiation.

All material properties density, thermal conductivity, specific heat, and surface emissivity were assigned based on literature values for ZBLAN glass, silica, air and tungsten. These parameters are summarized in Table 1.

*Table 1: Material properties and their corresponding values.*

| Properties/ Materials | Silica | ZBLAN | Air | Tungsten |
|---|---|---|---|---|
| Heat capacity at constant pressure(J/Kg.K) | 745 [8] | 300 [8] | 1004 [8] | 134 [8] |
| Density (Kg/m$^3$) | 2203 [9] | 4300 [9] | 1.29 [8] | 19300 [8] |
| Thermal Conductivity (W/mK) | 1.38 [9] | 0.628 [9] | 0.025 [8] | 196.65 [8] |
| Surface emissivity | 0.93 [10] | 0.31 [10] | - | 0.5 [10] |

**2.3 Simulation Parameters and Meshing**

The outer boundary of the heating layer was maintained at a constant 400 °C (673.15 K), while all internal domains were initialized at room temperature of 25 °C (298.15 K). Because the ampoule was fully evacuated, convective heat transfer was neglected, leaving conduction and surface-to-surface radiation as the only active mechanisms. In Design I, the ZBLAN received heat through both conduction (via direct contact with the silica wall) and radiation, whereas in Design II, only radiative transfer occurred due to the absence of contact.

The simulation recorded temperature evolution within the ZBLAN as a function of time, allowing assessment of both the heating rate and the time required to reach 400 °C under each configuration. A mapped quadrilateral mesh was applied, with refinement at the ZBLAN-silica interfaces and across the vacuum region to resolve steep thermal gradients. Mesh independence was confirmed when further refinement resulted in less than 1% variation in the maximum ZBLAN temperature.

This approach enables direct comparison of the thermal response between full-contact and contact-loss scenarios, quantifying how vibration-induced displacement affects heating efficiency and validating the hypothesis proposed in Part I.

### 3. Results and Discussion

### 3.1 Initial Thermal Calibration and Baseline Temperature Distribution

Before initiating the transient heat transfer simulations, the model was calibrated to verify thermal uniformity and ensure correct boundary implementation. Figure 2 presents the temperature distribution across the entire domain at the initial time step (0 minutes), immediately after the ampoule containing the ZBLAN sample was introduced into the pre-heated source maintained at 673.15 K (400 °C). At this moment, the ampoule and ZBLAN remain at ambient temperature (298.15 K), while the surrounding heater region exhibits a uniform high temperature.

As seen in the Figure 2, a distinct temperature contrast exists between the heating source and the ampoule. However, no thermal gradient is observed within the ampoule itself, confirming that heat transfer has not yet begun at this stage. Since the model is under vacuum conditions, convection is absent, and radiative and conductive mechanisms have not yet been initiated. The observed uniform temperature distribution inside the ampoule therefore represents the initial steady-state condition prior to the onset of heat exchange.

This calibration step confirms that the model correctly reproduces the experimental starting condition, with the ampoule at room temperature and the heater at a defined boundary temperature, providing a reliable baseline for subsequent transient analyses.

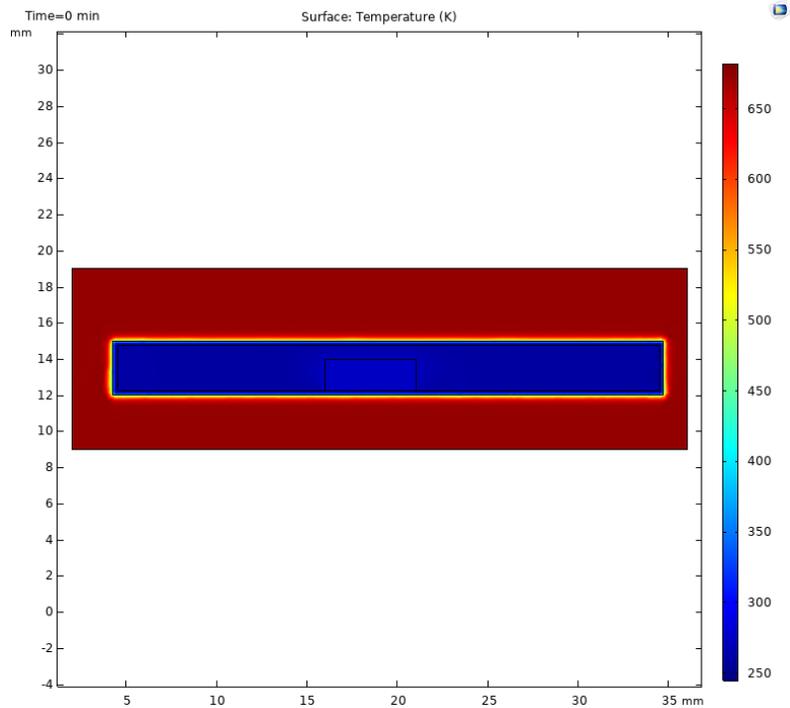

*Figure 2: Temperature distribution across the entire model domain at 0 minutes.*

Figure 3 presents COMSOL Multiphysics simulations of heat transfer in ZBLAN when a uniform temperature of 673.15 K is applied to all four walls of the heating zone for 1 minute for both design I and II.

Figure 3(a) illustrates heat transfer to the ZBLAN sample while in direct contact with the silica ampoule for a Design I. After 1 minute with the walls maintained at 673.15 K, the sample temperature spans 471- 494 K, with the highest values concentrated at the two lower corners. This localization is the classic corner effect arising from enhanced solid- solid conduction at geometric junctions.

Similarly, Figure 3(b) depicts heat transfer to the ZBLAN sample suspended freely inside the silica ampoule (no wall contact), based on Design II. Under the same boundary condition 673.15 K applied uniformly to all four walls for 1 minute the maximum temperature is 274 K, and the warmest zones appear at all four corners. Here the temperature pattern is governed primarily by radiative exchange: the corners experience relatively larger view factors and favorable geometric alignment, concentrating radiative energy and yielding a more symmetric four-corner hot-spot distribution.

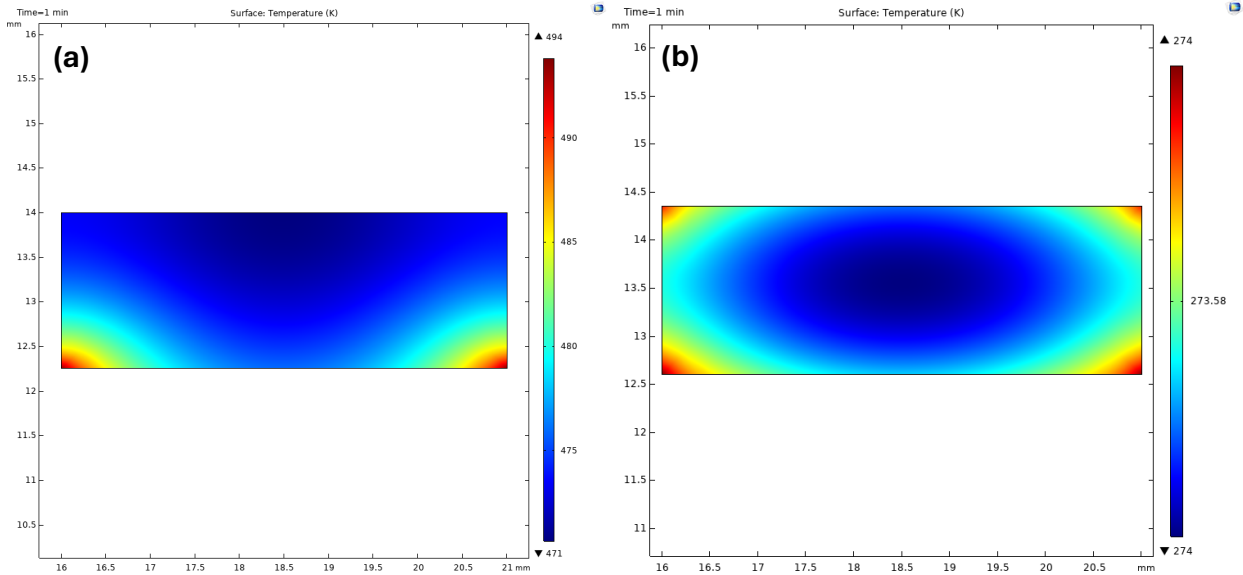

*Figure 3: Heat Flow in ZBLAN at 673.15 K (400 °C) after 1 minute of heating, simulated in COMSOL for (a) Design I, and (b) Design II.*

The corner effect observed in Figure 3 is illustrated in greater detail in Figure 4. The results correspond to the ZBLAN sample treated for one minute under Design I conditions. The conductive heat flux streamlines (black arrows) depict the direction of thermal energy transfer, revealing a pronounced concentration along the lower corners of the sample. These regions act as primary channels for heat conduction from the silica ampoule into the ZBLAN. After one minute of heating, the core temperature stabilizes at approximately 520 K. Within the surrounding vacuum region, a distinct radial thermal gradient develops, with successive layers exhibiting temperature increments of about 20 K. This layered temperature profile highlights the dominance of conduction in the absence of convection, as the vacuum minimizes thermal losses and enhances diffusion-driven heat transport. The localized accumulation of heat at the lower edges results in higher temperature intensity at these points, thereby producing the characteristic corner effect.

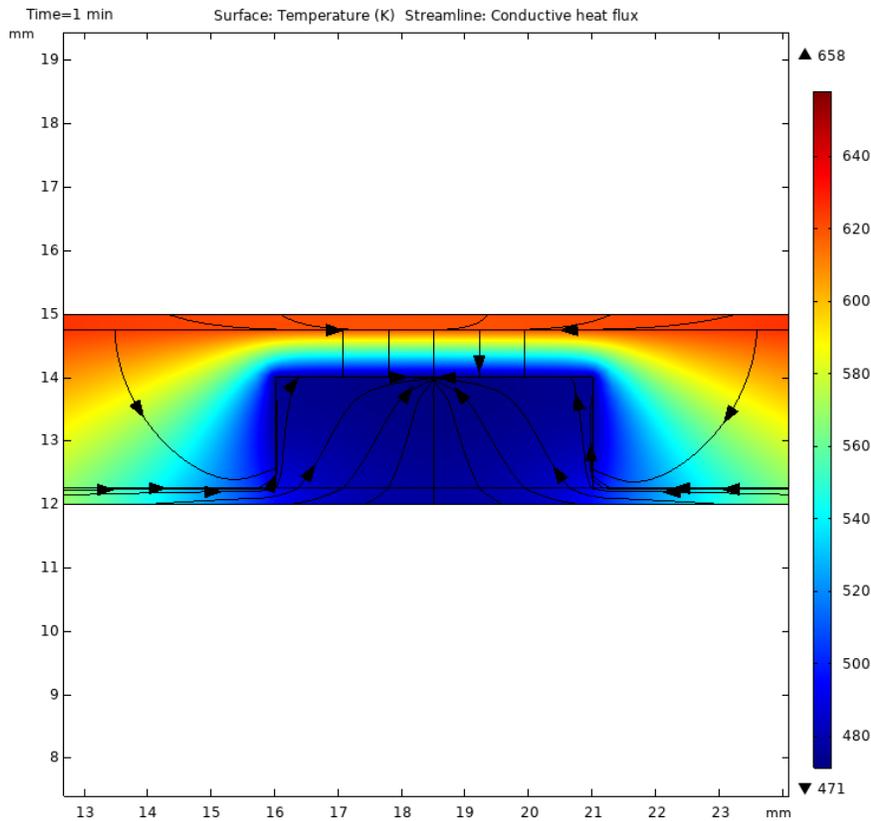

*Figure 4: Temperature distribution across the ampoule at 1 minute, highlighting the heat flux direction from the heating source to the ZBLAN sample in the Design I COMSOL simulation.*

Figure 5 presents the comparative study of the early-time heating response of ZBLAN under two COMSOL configurations over a 3-minutes interval, with all four heating-zone walls held at 673.15 K. In Design I, the curve exhibits a classic transient heating behavior: a steep rise during the first 1-2 minutes, followed by a progressive reduction in slope as the sample approaches its thermal steady state near 673 K (≈ 400 °C). By t ≈ 3 minutes, the temperature has climbed to roughly ~600 K, indicating that the system is converging toward equilibrium but has not yet fully saturated. In contrast, Design II shows a much more gradual, nearly linear increase, rising only from about 273 K to ~275-276 K over the same period. This pronounced disparity reflects the dominant solid-solid conduction in Design I, which enables rapid heat uptake, versus the radiation-limited heat transfer in Design II, which results in a far smaller heat-transfer coefficient and a correspondingly slower temperature rise. The figure thus highlights how boundary condition and contact mechanics govern early-time heating rates: Design I rapidly approach the imposed wall temperature, whereas Design II remains far from equilibrium within the 3 minutes observation window.

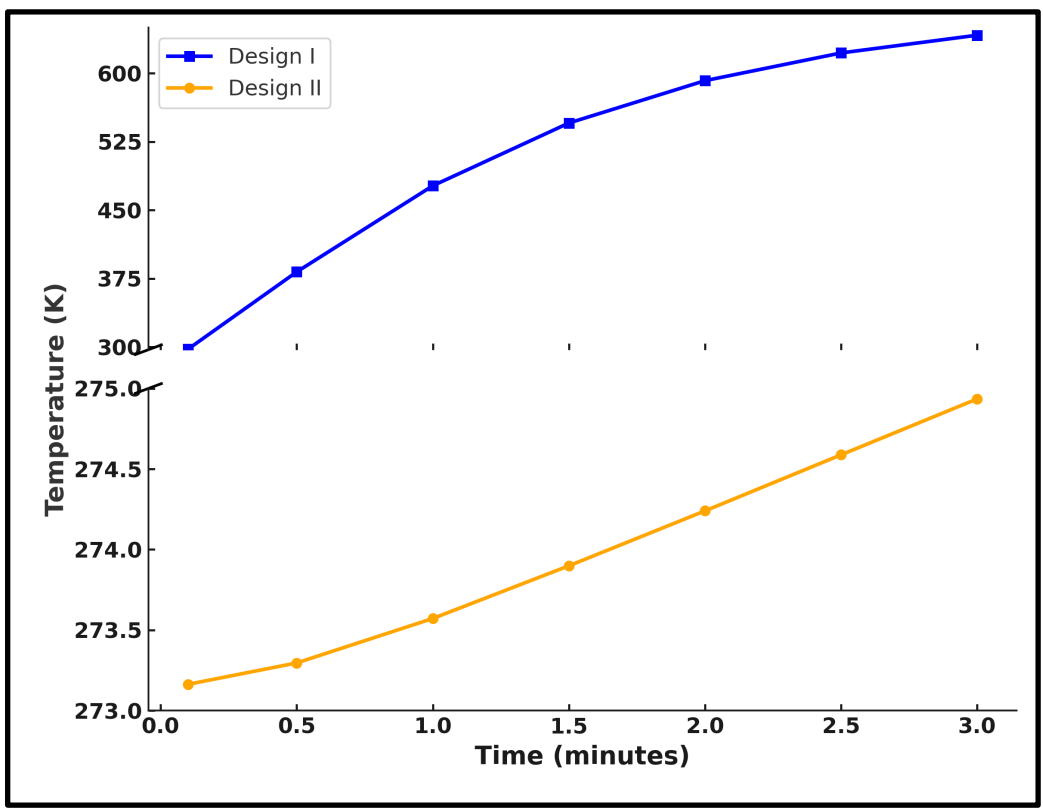

*Figure 5: Time-temperature curves of ZBLAN from COMSOL simulations for Design I and Design II.*

Overall, the COMSOL simulation results establish a clear quantitative relationship between thermal coupling and the heating behavior of ZBLAN. In Design I (contact configuration), the sample exhibits efficient energy uptake and a higher steady-state temperature, consistent with the corner effect where heat conduction intensifies at geometric junctions. In contrast, Design II (free-floating configuration) relies primarily on radiative transfer, resulting in a slower, nearly linear temperature increase and an overall cooler thermal profile. Combined with the spatial temperature fields, the time-temperature evolution confirms the hypothesis proposed in Part I: mechanical vibration that intermittently disrupts thermal contact suppresses conductive heat transfer, maintaining the material below the critical temperature required for crystallization. Hence, the variability observed in prior experiments arises from inadequate thermal contact rather than insufficient external heating.

**3.2 Apparatus Redesign and Experimental Validation**

With the COMSOL simulations confirming the proposed hypothesis, the experimental setup was refined to minimize the jostling effect observed during the vibration-assisted heat treatment described in Part I [5]. The redesigned configuration ensured continuous thermal contact between the ZBLAN sample and the inner wall of the silica ampoule throughout the test. To achieve this, the apparatus from the previous study was modified by elevating one side by approximately 4°, using concrete supports as shown in Figure 6. A layer of carpet was placed above the supports to replicate the non-inclined condition when required. All components including the vibration motor, clamping system, and heating unit were firmly secured with bolts and straps to prevent movement or mechanical interference during operation. Preliminary vibration-only trials verified that the inclination effectively restricted ZBLAN displacement, maintaining the sample at the lower edge of the ampoule, where stable heating through conduction and radiation could occur. To confirm that the vibrational environment remained consistent with the previous configuration, a secondary vibration analysis was conducted using the same tri-axial accelerometer setup employed in Part I [5]. The measured frequencies, shown in Figure 7, indicated that the redesigned apparatus preserved identical vibrational characteristics while successfully eliminating unintended sample movement.

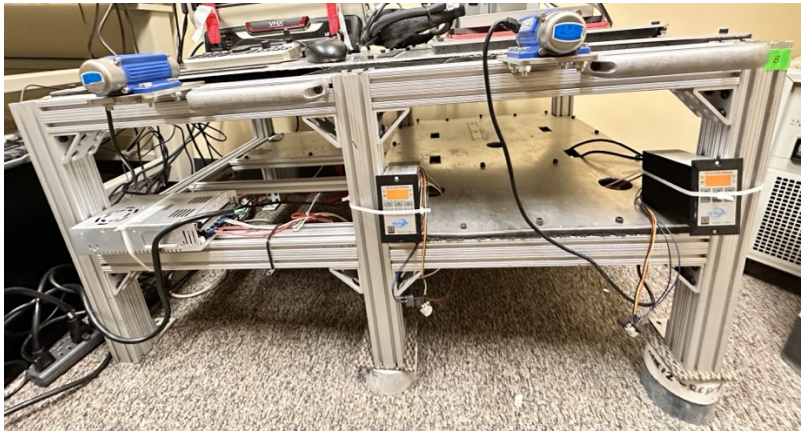

*Figure 6: ZBLAN vibration testing experimental set up with the inclination of 4º.*

To verify that the redesigned inclined setup preserved comparable vibration characteristics to the original configuration, vibration frequency measurements were collected for the high-speed vibrating motor using a tri-axial accelerometer. The overall vibration magnitude incorporating the contributions of all three orthogonal directions using vector sum [11] was calculated as,

$$Magnitude = \sqrt{X^2 + Y^2 + Z^2}$$

The results, shown in Figure 7, compare the vibration magnitude of the inclined and non-inclined setups across levels H1 to H5. At lower vibration levels (H1 and H2), the inclined setup produced slightly lower magnitudes, approximately 5 to 8 Hz below the non-inclined configuration. However, at higher vibration levels (H3, H4, and H5), both setups exhibited nearly identical vibration magnitudes, with differences remaining within 2 to 3 Hz. This close agreement indicates that the inclination modification, while eliminating sample jostling, did not alter the mechanical excitation at the higher operating frequencies.

Given that ZBLAN samples exhibited anomalous behavior primarily at vibration levels H3 to H5 in Part I, these levels were selected for retesting using the improved setup. The similarity in vibration magnitude between the two configurations ensures that any observed changes in crystallization behavior can be confidently attributed to the enhanced thermal contact rather than to variations in vibrational intensity.

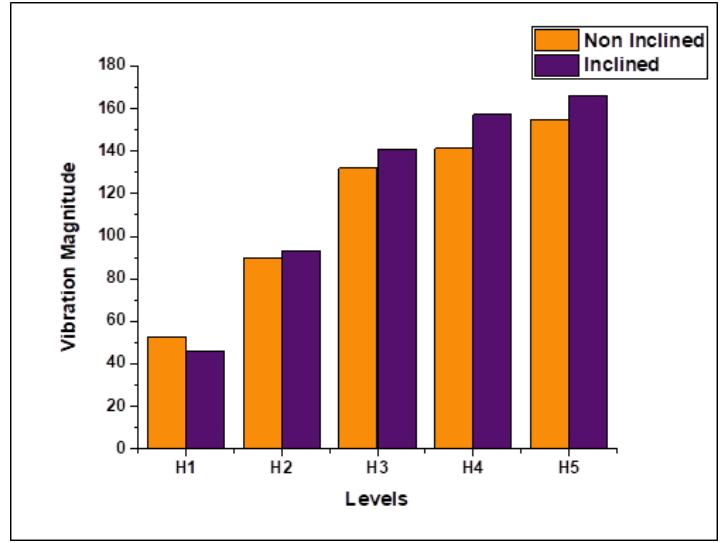

*Figure 7: Comparative vibrational frequency of non-inclined and inclined microscopic setups under high-frequency vibration.*

With the implementation of the inclined configuration, ZBLAN samples were subjected to higher vibration levels ranging from H3 to H5. To validate the simulation outcomes, the samples that had previously exhibited anomalous crystallization behavior under these vibration conditions were retested using the modified setup. This approach enabled direct comparison between the initial and improved configurations, isolating the influence of stable thermal contact on the crystallization process. The detailed comparative results, along with the subsequent microstructural and compositional characterizations, are presented in the following sections.

### 3.3 Microscopic Characterization

Comparative microscopic analyses of ZBLAN samples subjected to varying thermal and vibrational conditions are presented in Figure 8. The experiments were conducted using two configurations: the non-inclined setup described in Part I and the improved inclined setup developed in this study.

The inclined configuration produced consistent and well-defined morphological evolution across all vibration levels. Incipient crystallization appeared at vibration level H3 within the temperature range of 330-350 °C and at H5 between 330-340 °C. At H5 and 350 °C, partial crystallization was observed, progressing into a highly crystalline structure between 360 and 380 °C. Similarly, at H4, well-developed crystalline domains emerged around 360 °C. These transitions were absent in samples treated using the non-inclined configuration, confirming that the previously observed inconsistencies originated from insufficient heating caused by intermittent thermal contact loss.

Overall, the results demonstrate that the inclined configuration ensures stable sample positioning and uniform heat transfer, enabling accurate determination of crystallization onset and morphological progression. These findings confirm that the redesigned apparatus effectively mitigates the jostling-induced thermal decoupling reported in Part I, thereby providing reproducible and controlled conditions for investigating vibration-assisted crystallization in ZBLAN.

| Condition | Non- Inclined | Inclined |
|---|---|---|
| H3_330 | 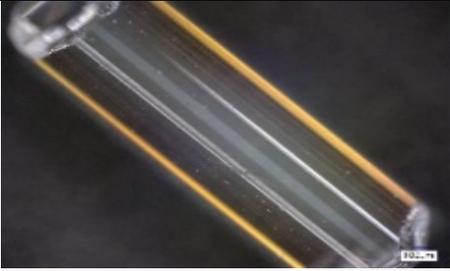 | 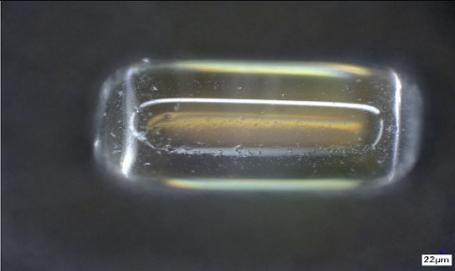 |
| H3_340 | 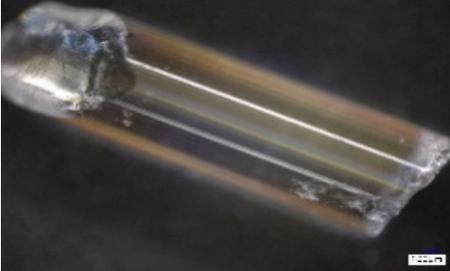 | 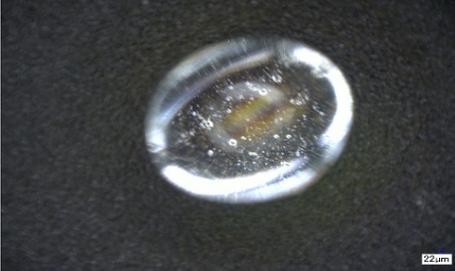 |
| H3_350 | 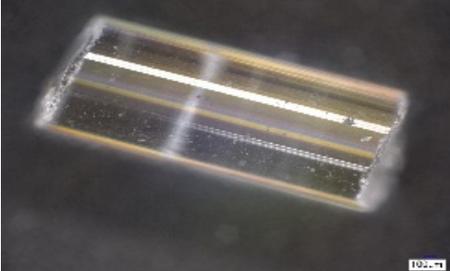 | 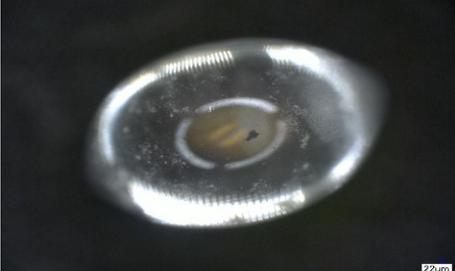 |
| H3_360 | 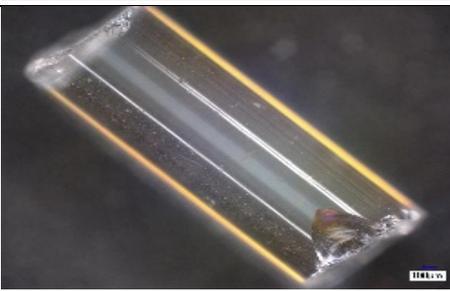 | 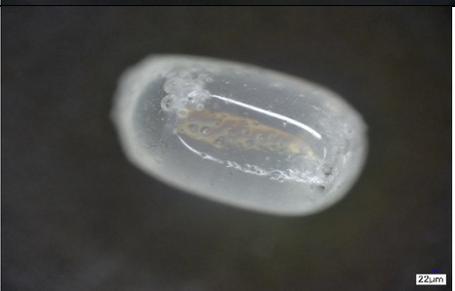 |

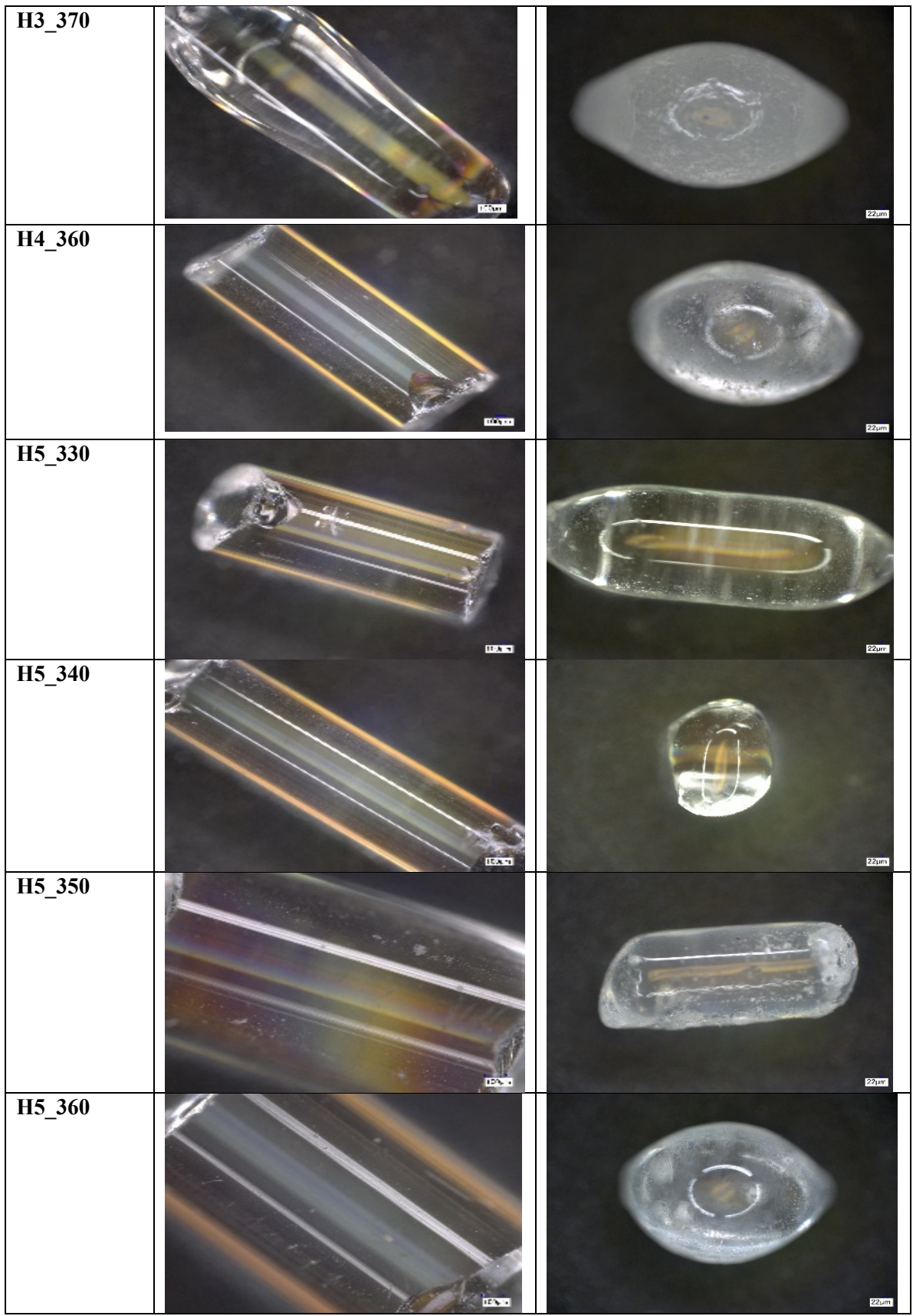

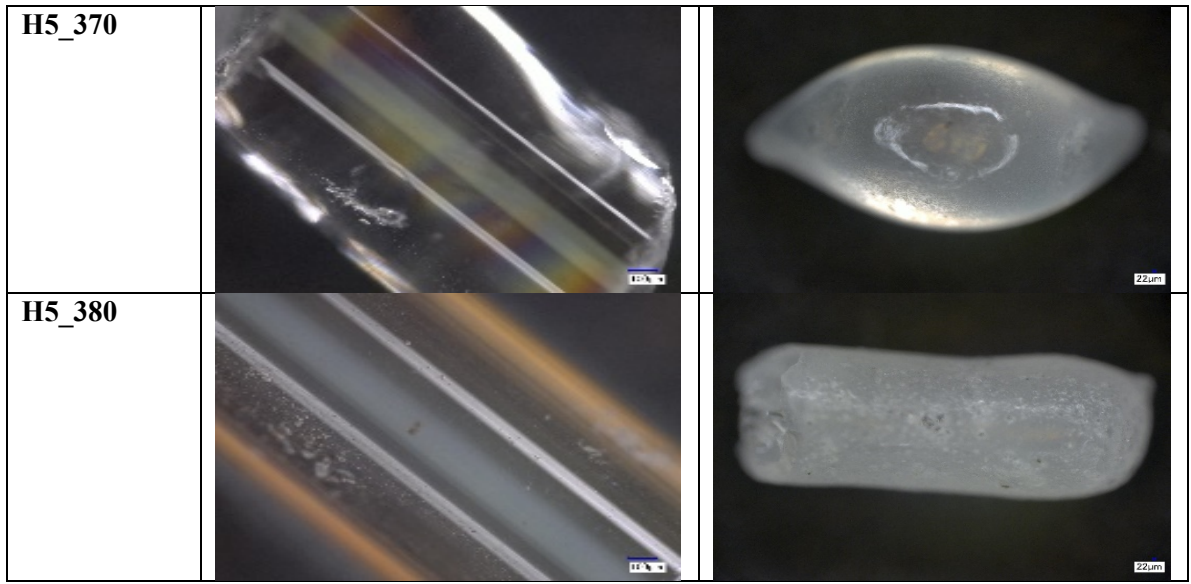

*Figure 8: Microscopic images of ZBLAN obtained using non-inclined and inclined experimental setups.*

### 3.4 Scanning Electron Microscopy (SEM)

To provide a broader understanding of crystallization behavior under combined thermal and vibrational conditions, additional SEM analyses were conducted on multiple ZBLAN samples treated across a wide range of vibration and temperature levels. Among these, selected SEM results are presented and discussed here, focusing on samples treated at all five vibration levels generated by the high-speed vibrating motor and at higher temperature conditions where pronounced crystal formation occurred. These representative samples enable detailed examination of the evolution in crystal morphology, including variations in crystal size, density, and shape, thereby offering deeper insight into the influence of vibration intensity and heating uniformity on ZBLAN crystallization.

At a temperature of 380 °C and a vibration frequency corresponding to level H1, extensive crystal formation was observed along the edges of the ZBLAN cladding layer as shown in Figure 9(a). The distribution of crystals along the boundary region indicates that the cladding acts as a preferred site for nucleation under combined thermal and vibrational excitation. Magnified regions from two distinct edge locations are presented in Figure 9(b) and Figure 9(c), each revealing unique morphological characteristics.

Figure 9(b) displays a prominent starburst pattern, where multiple bow-tie-shaped crystals radiate outward from a single nucleation center. This configuration reflects localized heterogeneous nucleation, likely initiated at a surface irregularity, compositional variation, or micro-defect acting as a preferred crystallization site. The radial symmetry of this structure suggests that once nucleation commenced, crystal fronts propagated rapidly outward under sustained heat and vibration, resulting in a well-defined, symmetrical starburst formation. The crystals in this region occupy approximately 24.44 % of the total SEM image area, corresponding to 6172.88 µm², with the largest individual crystal measuring 162.21 µm².

In contrast, Figure 9(c) reveals feather-like crystals coexisting with bowtie and dark needle-shaped morphologies extending along the edge of the ZBLAN surface shown in Figure 9(a). The crystals in this region cover 25.21 % of the SEM image, totaling 1877.02 µm², with the largest crystal measuring 47.98 µm². The coexistence of multiple morphologies across adjacent regions emphasizes the heterogeneous and directional nature of nucleation, indicating that vibration-assisted heat transfer enables simultaneous activation of multiple nucleation mechanisms and promotes anisotropic crystal growth along the cladding boundary.

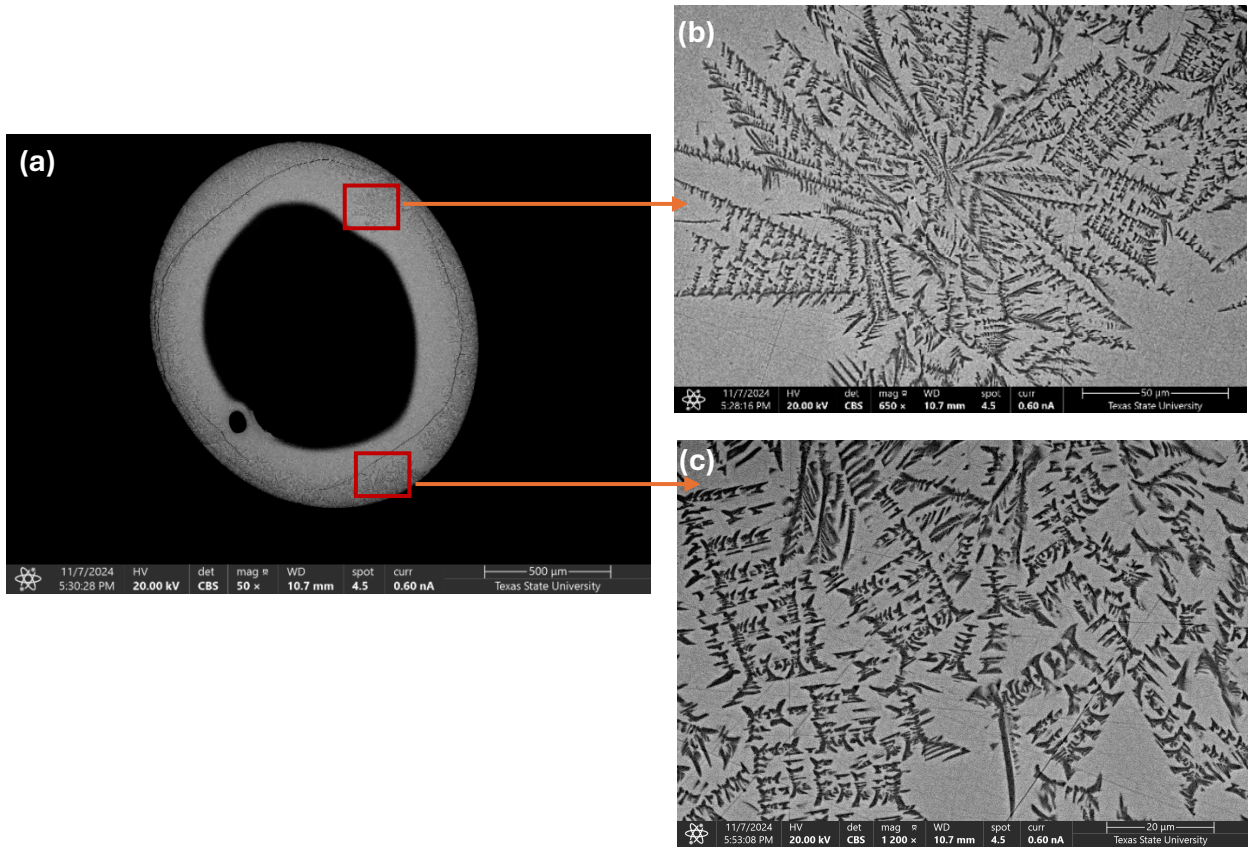

*Figure 9: SEM images of ZBLAN treated at a vibration level H1 and temperature of 380 °C (a) full view showing the core and cladding layers, (b) magnified upper region of the cladding layer, and (c) magnified lower region of the cladding layer.*

Distinct crystallization was also evident for the sample processed at level H2 and 370 °C as shown in Figure 10. The surface contained predominantly dark needle-shaped crystals intermixed with bowtie and feather morphologies. These features collectively covered 29.06 % of the image area, corresponding to a total crystal area of 7343.15 µm², and the largest crystal reached 297.84 µm². The emergence of well-developed crystals under slightly reduced temperature conditions suggests that mechanical vibration provided sufficient localized energy to enable nucleation and early crystal growth even near the thermal threshold.

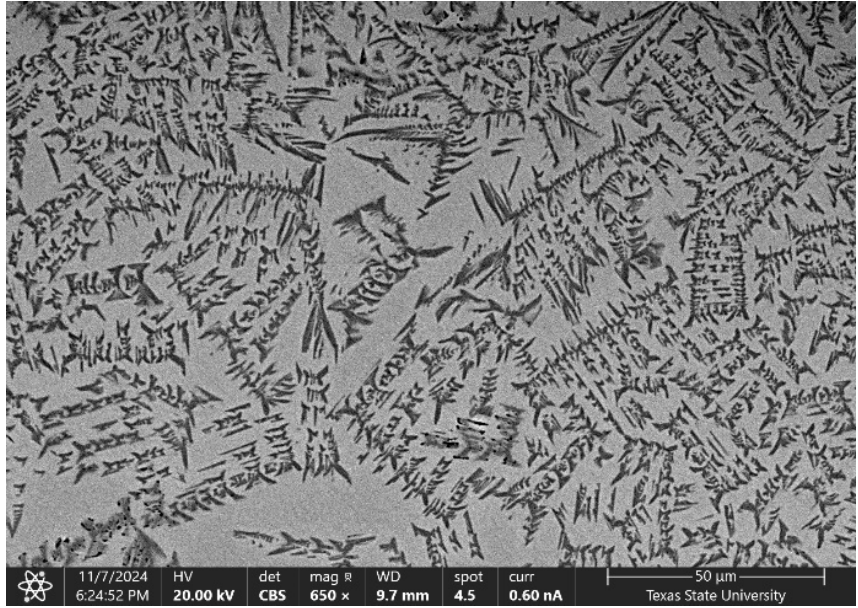

*Figure 10: SEM images of ZBLAN treated at a vibration level H2 and temperature of 370 °C, showing surface morphology and crystal formation features.*

Further increases in both vibration frequency and temperature, represented by the H3 sample treated at 390 °C, produced dense and broadly distributed crystallization across the ZBLAN surface as shown in Figure 11. The microstructure was dominated by elongated needle-shaped and bow-tie crystals, with a small fraction of feather-like structures. Crystals occupied 26.58 % of the image area (6767.99 µm²), and the largest measured 313.25 µm². The extensive coverage and uniform distribution indicate a transition from isolated nucleation to a continuous crystallization regime, reflecting the enhanced atomic diffusion and directional growth promoted by sustained vibration.

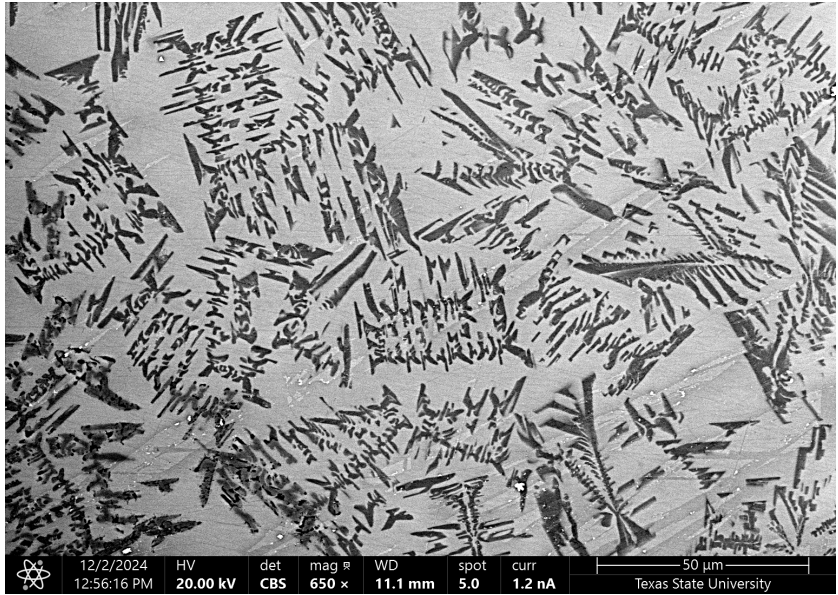

*Figure 11: SEM images of ZBLAN treated at vibration level H3 and temperature of 390 °C, showing the developed crystalline morphology.*

At level H4 and 380 °C, the surface exhibited a dense and interconnected crystalline network composed primarily of bowtie and feather-like crystals as shown in Figure 12. The crystalline fraction reached 26.98 %, corresponding to a total area of 2907.55 μm², and the largest crystal measured 122.66 μm². This morphology points to the establishment of stable, overlapping nucleation regions and grain coalescence driven by increased vibration intensity and uniform heating.

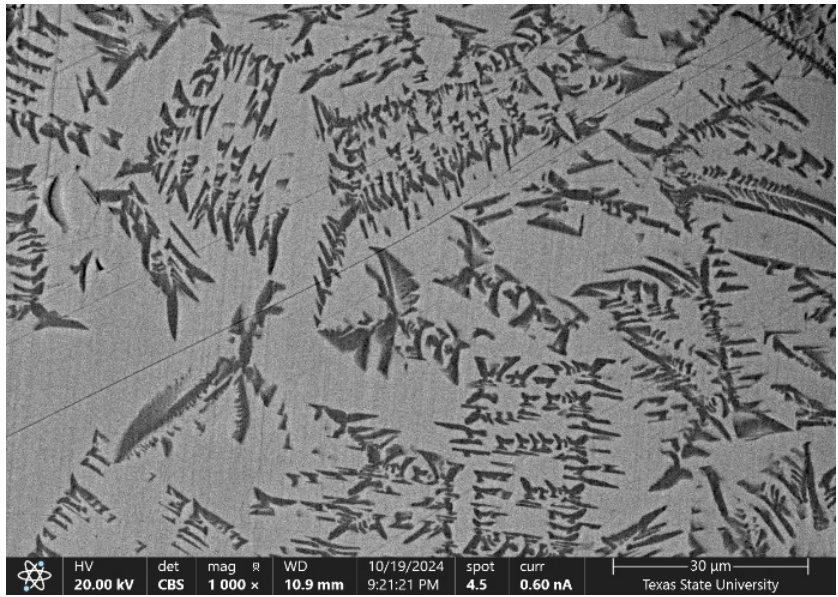

*Figure 12: SEM image of ZBLAN treated at vibration level H4 and temperatures of 380 °C.*

Similarly, the microstructure of the H5 sample treated at 380 °C revealed a fine and continuous distribution of crystallites covering the surface almost uniformly as shown in Figure 13. Quantitative analysis showed an average crystal size of 2.706 μm², with total crystalline coverage of 20.616 μm² and the largest crystal measuring 16.122 μm². These findings suggest that the highest vibration level enhances nucleation frequency but limits grain growth, producing a dense network of small, equiaxed crystals.

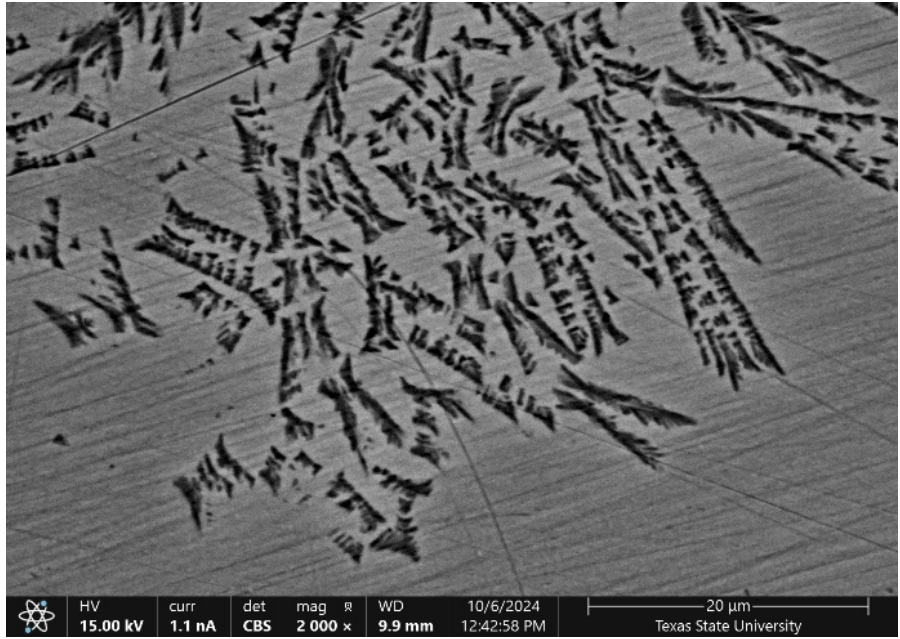

*Figure 13: SEM image of ZBLAN treated at vibration level H5 and temperature of 380 °C.*

At a vibration frequency of level H5 and a temperature of 400 °C, the ZBLAN surface exhibited pronounced crystallization features, as shown in Figure 14. In Figure 14(a), feather-like crystal morphologies dominate the surface, accompanied by several dark needle-shaped crystals dispersed along the cladding region. These structures collectively occupy 19.61 % of the total SEM image area (2109.36 μm²), with the largest crystal measuring 1010.11 μm². In contrast, Figure 14(b) reveals a mixed population of feather-shaped, bow-tie-shaped, and dark needle-shaped crystals. Here, the crystals cover 29.06 % of the total SEM image area (7434.15 μm²), with the largest crystal reaching 297.84 μm². The coexistence of multiple morphologies at this elevated temperature and high vibration level suggests that the strong mechanical excitation enhances atomic rearrangement and promotes diverse nucleation pathways, leading to concurrent growth of multiple crystalline forms across the ZBLAN surface.

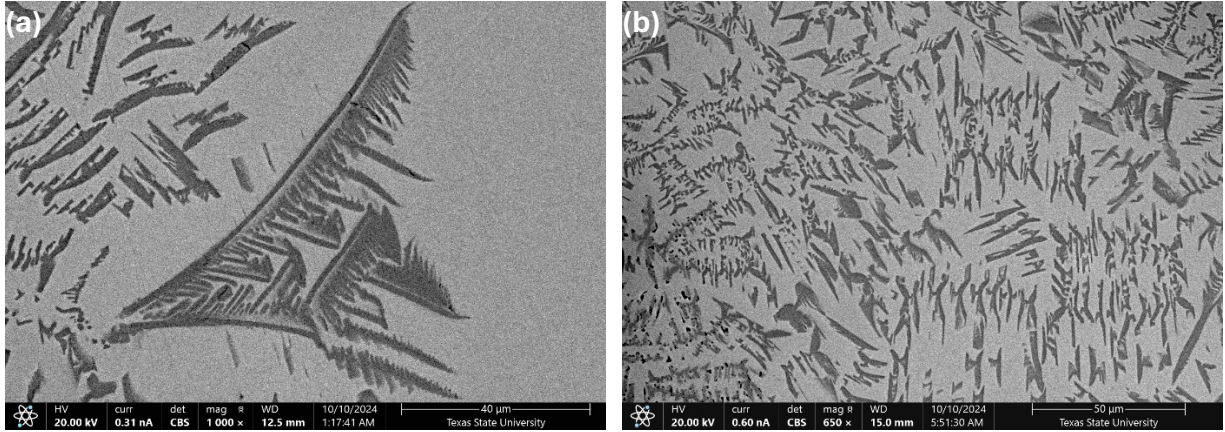

*Figure 14: SEM images of ZBLAN treated at vibration level H5 and temperature of 400 °C.*

Taken together, these SEM analyses reveal a consistent evolution in crystallization behavior with increasing vibration and temperature. At low vibration, crystallization initiates locally along the cladding boundary through heterogeneous nucleation. As vibration intensity increases, nucleation becomes more widespread and uniform, leading to dense networks of interconnected crystals and eventually fine-grained microstructures at the highest excitation levels. These results validate that strong mechanical vibration enhances atomic mobility, accelerates heat transfer, and governs the density and morphology of crystals formed in ZBLAN.

### 3.5 Energy Dispersive X-ray Spectroscopy (EDS)

A comparative energy dispersive x-ray spectroscopy (EDS) analysis was performed on the ZBLAN sample treated at a vibration of level L3 and a temperature of 380 °C, where several crystals with distinct morphologies were observed across the cladding region. Among these, two representative crystal types, the feather-shaped and the dark needle-shaped, were selected for compositional analysis to examine possible elemental variations associated with different morphologies. Line mapping was conducted with ten equally spaced points across each crystal to determine the corresponding elemental weight percentages and spatial distribution as shown in Figure 15.

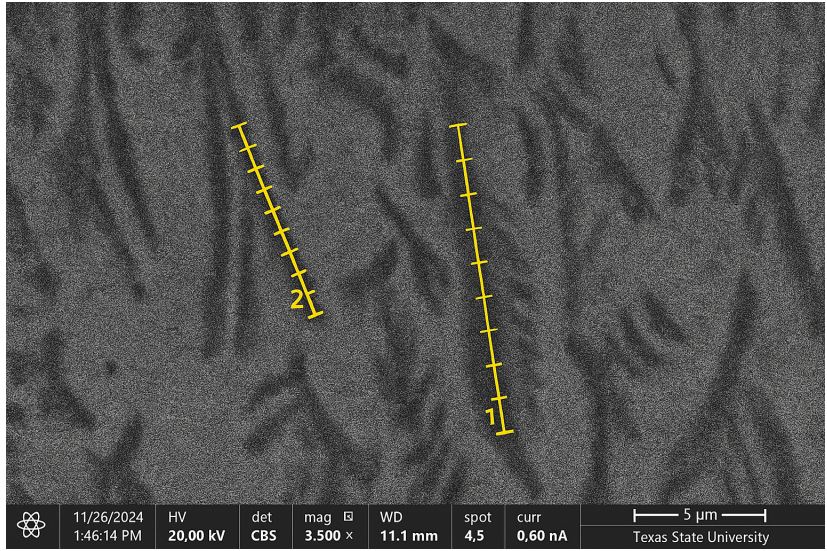

*Figure 15: ZBLAN showing feather-like and dark needle-shaped crystals after treatment at vibration level L3 and temperature of 380 °C.*

The feather-shaped crystal, labeled as region 1 in Figure 15, exhibited a composition primarily dominated by hafnium and fluorine. As shown in Figure 16, hafnium constituted approximately 50.72 wt % (± 1.06 %), followed by fluorine at 32.24 wt % (± 0.76 %). Barium and sodium were present in smaller quantities, below 10 wt %, with respective standard deviations of 1.52 % and 0.40 %. Aluminum appeared only in trace amounts, and zirconium was not detected, confirming that this crystal originated within the cladding layer rather than the Zr-rich core region.

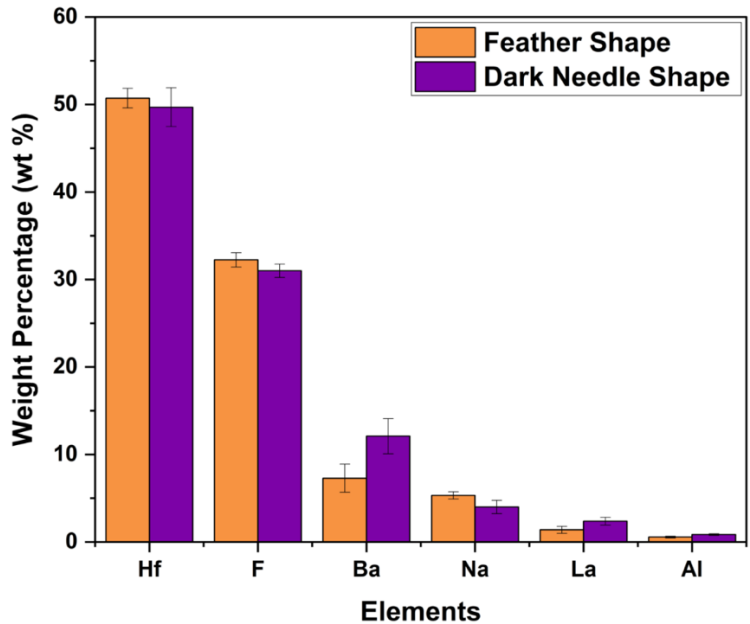

*Figure 16: Elements vs. weight percentage along the feather-like and dark needle-shaped crystals shown in Figure 15.*

The dark needle-shaped crystal, labeled as region 2 in Figure 15, displayed a similar composition pattern but with noticeable variations in secondary element concentrations. Hafnium remained the predominant element at approximately 50 wt % (± 2.1 %), while fluorine accounted for 31 wt % (± 0.7 %). Barium concentration increased to 12.09 wt % (± 1.91 %), and minor amounts of sodium and lanthanum were detected, each below 5 wt %. Aluminum was again negligible, and no zirconium was identified.

A direct comparison of the EDS spectra, as illustrated in Figure 16, indicates that both crystal morphologies are hafnium and fluorine rich phases typical of ZBLAN cladding crystallization. However, the higher barium concentration in the dark needle-shaped crystal suggests localized compositional segregation during solidification. These variations point to the influence of temperature uniformity and vibration amplitude on atomic diffusion and element partitioning, ultimately driving the formation of distinct crystal morphologies within the same material system.

4. **Conclusion**

The inconsistencies reported in Paper I, where ZBLAN samples exhibited unpredictable crystallization behavior at higher vibration levels, led to the hypothesis that intermittent jostling within the ampoule disrupted thermal contact and caused non-uniform heating. This hypothesis was verified in the present study through COMSOL Multiphysics simulations, which demonstrated that the loss of physical contact between the ZBLAN and the silica ampoule wall markedly reduced heat transfer efficiency. Incorporating a four-degree inclination into the experimental setup successfully restored stable thermal coupling, and subsequent experiments confirmed uniform crystallization, thereby validating the computational predictions.

*Table 2: Crystallization stages of ZBLAN under varying vibration frequencies and temperatures for inclined and non-inclined setups.*

| Vibrator/Level | Vibration Magnitude (Hz) | Temperature (ºC) | | | | | | | | | | |
|---|---|---|---|---|---|---|---|---|---|---|---|---|
| | | 250 | 300 | 320 | 330 | 340 | 350 | 360 | 370 | 380 | 390 | 400 |
| | 0 (Control) | green | green | green | green | yellow | yellow | orange | orange | red | red | red |
| High - One | 52 | green | green | green | green | yellow | yellow | red | red | red | red | red |
| High - Two | 90 | green | green | green | yellow | yellow | yellow | red | red | red | red | red |
| High - Three | 130 | green | green | green | yellow | yellow | yellow | red | red | red | red | red |
| High - Four | 140 | green | green | green | yellow | yellow | orange | red | red | red | red | red |
| High - Five | 170 | green | green | green | yellow | yellow | orange | red | red | red | red | red |
| Low - Two | 65 | green | green | green | yellow | yellow | yellow | red | red | red | red | red |
| Low - Three | 80 | green | green | green | yellow | yellow | yellow | red | red | red | red | red |
| Low - Four | 90 | green | green | green | yellow | yellow | yellow | red | red | red | red | red |

Building upon the microscopic, SEM, EDS, and AFM analyses from Paper I and the current study, a unified color-coding framework was developed to classify the crystallization behavior of ZBLAN samples, as summarized in Table 2. In this scheme, green represents amorphous glass, yellow denotes incipient crystallization, orange corresponds to partial crystallization, and red indicates a highly crystallized state. Representative microscopic images and corresponding structural descriptions are provided in Table 3. This integrated classification system offers a comprehensive overview of ZBLAN's structural evolution under combined thermal and vibrational influences, facilitating clearer interpretation of crystallization dynamics across varying processing conditions.

*Table 3: Color-coded ZBLAN structures with representative microscopic images.*

| Sample Structure | Color Code | Representative Images |
|---|---|---|
| Amorphous | 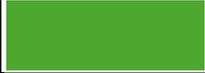 | 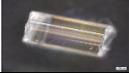 |
| Incipiently Crystallized | 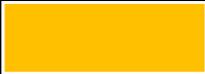 | 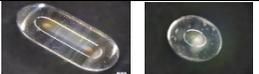 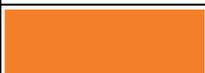 |
| Partially Crystallized | 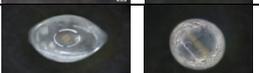 | 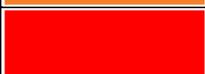 |
| Highly Crystallized | 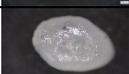 | |

The classification results reveal that under purely thermal treatment, amorphous ZBLAN remained stable up to approximately 340 °C and gradually transitioned into a partially crystallized state between 360 °C and 370 °C. When low level vibration was introduced, crystallization commenced earlier around 330 °C with full transformation achieved by 360 °C. At higher vibration frequencies (above 100 Hz), complete crystallization initiated near 350 °C, indicating that excessive vibration is unsuitable for fiber drawing. Elevated vibration levels accelerated nucleation and produced an intermediate partially crystallized phase that was absent at lower frequencies. These findings suggest that vibration enhances atomic mobility and facilitates phase transformation by improving localized heat transfer and promoting multiple nucleation pathways.

Overall, the combined computational and experimental results demonstrate that both thermal and vibrational parameters critically influence the crystallization kinetics of ZBLAN. Sustained exposure above 330 °C under vibration significantly increases the risk of crystallization, effectively narrowing the viable fiber drawing window by approximately 30 °C. For microgravity-based processing environments such as aboard the International Space Station, maintaining the amorphous structure of ZBLAN requires precise control of both temperature and vibration. Effective damping of mechanical excitation at elevated temperatures or complete isolation during heating is essential to prevent phase instability. This study establishes a predictive framework for designing vibration resistant processing protocols to enable the reliable fabrication of optically pure, crystal free ZBLAN fibers.